\documentclass[aps,pre]{revtex4}

\usepackage{graphicx}
\usepackage[utf8]{inputenc}
\usepackage[english]{babel}
\usepackage{amsmath}
\usepackage{graphicx,enumerate}
\usepackage{amssymb}
\begin{document}

\title{Physical interpretation of the canonical ensemble for long-range interacting 
systems in the absence of ensemble equivalence}

\author{Marco Baldovin}
\affiliation{Dipartimento di Fisica, Universit\`a di Roma Sapienza, P.le Aldo Moro 2, 00185, Rome, Italy}

\date{\today}

\begin{abstract}
In systems with long-range interactions, since energy is a non-additive quantity,
ensemble inequivalence can arise: it is possible
that different statistical ensembles lead to different equilibrium descriptions,
even in the thermodynamic limit. The microcanonical ensemble should be considered  
the physically correct equilibrium distribution as long as the system is isolated.
The canonical ensemble, on the other hand, can always be defined mathematically, but it is quite natural
to wonder to which physical situations it does correspond.
We show numerically and, in some cases, analytically, that the equilibrium properties
of a generalized Hamiltonian mean-field model in which ensemble inequivalence
is present are correctly described by the canonical distribution in (at least) two different
scenarios: a) when the system is coupled via local interactions to a large reservoir 
(even if the reservoir shows, in turn, ensemble inequivalence) and b) when the mean-field
interaction between a small part of a system and the rest of it is weakened by
some kind of screening.
\end{abstract}

\maketitle

\section{Introduction}

Equilibrium statistical mechanics provides a very accurate description of the
statistical features of systems with many particles. Relevant results
can be derived when only short-range interactions are involved and
the thermodynamic limit is considered; among them, equivalence of statistical ensembles
covers a prominent role, since it allows the computation of averages for macroscopic 
observables according to different statistical descriptions \cite{huang1988}.
From a technical point of view it relies on the validity
of the law of  large numbers and of the central limit theorem, on the results 
of large deviations theory, but also on the concavity of thermodynamic
potentials \cite{touchette09}.
More difficult cases are:
\begin{itemize}
 \item systems  at the critical point where also spatially far parts
are strongly interacting, so that the central limit theorem  cannot be used (see
e.g. \cite{ma85, sethna05});
\item systems with few degrees of freedom \cite{seifert12, cerino15, puglisi17};
\item systems with long-range interactions, in which potentials
decay not faster than $r^{-d}$, where $r$ is the distance and $d$ the spatial
dimension \cite{dauxois02}.
\end{itemize}
The latter case includes rather interesting physical problems,
e.g. in plasma, hydrodynamics, self gravitating systems and 
lasers \cite{dauxois02}. In addition, all systems in which the elements interact via a mean field
also belong to this category.

In systems with long-range interactions the equivalence of
statistical ensembles is not guaranteed: in particular there are 
rigorous results for Hamiltonian models with mean field interactions,
showing that the thermodynamic potentials can be non convex; this is due 
to the non-additivity of energy \cite{campa09}.,graphicx,enumerate
As a consequence, the canonical and microcanonical
ensembles can give different results, i.e. the average
of a macroscopic observable $A$ is sensitive to the choice of the probability
density function:
$$
\langle A \rangle_m \neq \langle A \rangle_c \,.
$$
In other words, fixing the energy $E$ of a system does
not always lead to the same average one gets by fixing its temperature to the corresponding value
$T(E)\equiv \left(\frac{\partial S}{\partial E}\right)^{-1}$, where $S$ is the microcanonical
entropy. These  results are rather clear from a mathematical point of view, but their
physical meaning may appear not completely obvious, due to some potential sources of
confusion in the ``operative definition'' of the canonical ensemble for long-range
interacting systems.

Microcanonical ensemble always possess a transparent physical interpretation,
since it describes the statistical properties of isolated Hamiltonian systems.
The canonical ensemble, on the other hand, should be used for systems at fixed
temperature; it characterises, in particular, systems of Brownian particles, 
where the stochastic forces and the dissipation provide a constraint on the
temperature: such mechanism usually originates from the interactions of the particles
with another system (of a different nature) which acts as a stochastic thermal bath.
The above discussion is valid regardless of the range of the potential, and both
microcanonical and canonical ensemble have been extensively studied also for 
systems showing long-range interactions \cite{chavanis06}. Clearly, every Hamiltonian system (which is
described by the microcanonical ensemble as long as it is isolated) can be related
to a Brownian system (which is instead correctly described by the canonical ensemble):
notable examples are the relation between stellar systems and self-gravitating Brownian
particles \cite{sire04} and that between the Hamiltonian Mean Field model \cite{antoni95} and
the Brownian Mean Field model \cite{chavanis14}. 

As far as Hamiltonian systems with only short-range interactions are considered, the canonical ensemble
can be defined in a different way: it is generally possible to observe the statistical behaviour of a 
small number of degrees of freedom and regard the rest of the system as a thermal bath
constraining the temperature of such small portion. The procedure can be found on textbooks~\cite{huang1988}
and requires that the Hamiltonian term which represents the reciprocal interaction
is negligible in the thermodynamic limit. In this case the temperature is
fixed in a natural way, even in absence of an ``external'' reservoir.
As soon as long-range interactions are involved, the above procedure
cannot be applied: ``surface contributions'' to the energy of the small part,
due to the interactions with the rest of the system, are no more negligible
(i.e. energy is a non-additive quantity) and canonical ensemble cannot be defined 
in this way.

In past years some authors claimed that systems with long-range interactions
should be only described by the microcanonical ensemble~\cite{gross02, gross_campa02}.
It has also been pointed out that for self-gravitating systems canonical ensemble could
be only defined at a formal level~\cite{padmanabhan90}.
In the light of the above, other people stressed instead the role of canonical ensemble
in describing systems of Brownian particles coupled to external baths~\cite{chavanis_campa02}.
Operative protocols have also been studied in order to model a ``physical'' thermal
reservoir in numerical simulations, and their effects on the system have been
compared to those of Nos\'{e}-Hoover thermostats and Monte-Carlo integration
schemes in non-equilibrium conditions \cite{baldovin06, baldovin07, baldovin09}.

In this paper we address the problem of the physical meaning of canonical
ensemble when mean-field interacting systems with non-equivalence of ensembles
are involved; in particular, we show by numerical simulations that the canonical
ensemble is the only one that provides the correct equilibrium behavior
\begin{itemize}
 \item when the system is coupled via small local interactions to a large thermal
 bath;
 \item when the (mean-field) interaction between a small part of the system and
 the rest of it is very weak.
\end{itemize}
In the following we will study the Generalized Hamiltonian
Mean Field (GHMF) model introduced in Ref.~\cite{debuyl05}. This system is a generalization
of the well-known Hamiltonian Mean Field model \cite{antoni95}; it is composed of $N$
rotators whose Hamiltonian (with an additive constant) is:
\begin{equation}\label{ham}
 H_N=\sum_{i=1}^N \frac{p_i^2}{2} +N \left[ \frac{J}{2}(1-m^2) +\frac{K}{4}(1-m^4) \right]
\end{equation}
where $J$ and $K$ are constant parameters, $m$ is the intensity of a magnetization defined as
\begin{equation}\label{magn}
m=\sqrt{m_x^2+ m_y^2}\quad\quad m_x=\frac{1}{N}\sum_{i=1}^N \sin\theta_i \quad\quad m_y=\frac{1}{N}\sum_{i=1}^N \cos\theta_i\,
\end{equation}
and $\{\theta_i, p_i\}\quad i=1,...,N$ are canonical variables.
The statistical properties of GHMF model can be analytically studied using large-deviations
techniques \cite{campa09}. This approach shows that an isolated system can be characterized by 
negative specific heat $\partial \varepsilon/\partial T <0$ (where $\varepsilon$ is
the specific energy and $T$ the system's temperature) in a certain energy range for
suitable choices of $J$ and $K$. Therefore, microcanonical and canonical
ensembles are not equivalent, so that the graph of $T(\varepsilon)$ in the latter
description is not the inverse of $\varepsilon(T)$ in the former (it is necessary
to introduce a Maxwell construction, since a first order phase transition occurs in
the canonical ensemble).

The paper is organized as follows.
Section \ref{sec:baths} is devoted to the investigation of different
protocols to build a ``physical'' thermal reservoir for the GHMF model.
We show by numerical simulations that when the system is
coupled to the thermal bath by local interactions, its thermodynamic behavior is
described by the canonical ensemble, and ensemble inequivalence is clearly evident;
this is also true in the not completely trivial case in which the reservoir is a
GHMF system as well (therefore exhibiting negative specific heat). In Section 
\ref{sec:weak} the related problem of the equilibrium properties of a weakly
interacting portion of a GHMF system is investigated. We introduce a parameter
$\lambda$ which tunes the mean-field interaction between two portions
of the system: $\lambda$ determines how much each of the two subsystems ``feels'' the
mean-field effect of the other, varying between 0 (two isolated GHMF systems) and 1
(a unique GHMF system resulting from the complete mean-field interaction of the
two parts). The equilibrium behavior of a small portion of the system as a function
of $\lambda$ is analyzed using large deviation theory and molecular dynamics simulations:
in the $\lambda\ll 1$ limit, the canonical description is recovered.

In Section \ref{sec:conclusions} we briefly sketch our final remarks.

\section{Locally coupled ``thermal baths'' for systems with non-equivalence of ensembles}
\label{sec:baths}
In the following we consider three different ways of building a ``thermal
reservoir'' in numerical simulations. Each reservoir is coupled to a small
GHMF system \eqref{ham} with $J=1$, $K=10$. It has been shown \cite{patelli_vulpiani14} that this
choice of the parameters leads to first-order phase transitions in both microcanonical
and canonical ensembles; the latter is a direct consequence of the non-equivalence.

In this Section we consider ``local'' couplings: each particle of the system interacts
with only one particle of the bath. The coupling potential is given by an
Hamiltonian term $\lambda V_{coup}(\delta)$, where $\lambda$ is a (small) constant
which indicates the strength of the interaction
and $V_{coup}$ is a function of the angular distance $\delta$  between the two particles.
We choose:
 \begin{equation}\label{vcoup}
  V_{coup}(\delta)= A-B\cos\delta-C\cos^2\delta
 \end{equation}
with $A=J/2+3K/8$, $B=J/2+K/4$ and $C=K/8$, which is the interaction term of
Hamiltonian \eqref{ham} when $N=2$. There is no particular reason to make 
this choice for $V_{coup}(\delta)$, and the results should be quite independent of
its form, provided that its contribution to the total Hamiltonian is negligible.

Unless otherwise specified, molecular dynamics simulations reported in the present
and in the following Section are performed using a second-order Velocity
Verlet scheme, in which we take time steps short enough to get energy 
fluctuations of order $O(\Delta E/E)\approx 10^{-5}$.
Since we are interested in long-range interacting models at equilibrium, we compute
averages, as far as we can, \textit{after} thermalization, i.e. after the
system has departed from possible metastable states. Such process can take very
long times, depending on the total number of particles $N$ (see Ref.~\cite{yamaguchi04, debuyl11, pakter11}):
for this reason, in our simulations we choose relatively small values of $N$ (but still in the limit
$N\gg1$), namely $N\approx O(10^2)$ for the system and $N_{res}\approx O(10^3)$ for
the reservoir.
Initial values for positions and momenta are chosen according to Gaussian distributions, and
then rescaled in order to get the needed total energy; however we stress that,
since averages are computed after long thermalization times, our equilibrium results
should hold independently of the particular choice of initial conditions.

\subsection{Stochastic heat-bath}
First we study a bath composed of $N_{res}$ particles held at a fixed $T$ by a stochastic
term in its evolution equation: this term should model the effect of several
``collisions'' occurring on the rotators of the reservoir. We choose $N_{res}=N$, where $N$ is the
number of elements in the analyzed system, so that every particle of the system is coupled to exactly one
particle of such reservoir; consequently, the complete Langevin equation describing the motion of a single 
rotator in the bath (identified by an angular position $\xi_i$ and a momentum
$\pi_i$) reads:
\begin{equation}
\begin{cases}
\dot{\xi}_i=\pi_i\\
 \dot{\pi_i}= -\frac{1}{\tau}\pi_i + \sqrt{\frac{2T}{\tau}}\eta_i(t)
 -\lambda\frac{d}{d\xi_i} V_{coup}(\theta_i-\xi_i)
 \end{cases}
\end{equation}
where $\tau$ is a characteristic time of the system, $\eta_i(t)$ is a delta-correlated
Gaussian noise with zero mean such that $\langle\eta_i(t)\eta_j(t')\rangle=\delta_{ij}\delta(t-t')$
and $\theta_i$ is the angular position of the coupled particle in the system.
Here the Boltzmann's constant is 1.
On the other hand, the motion equations for a particle of the system are:
\begin{equation}
\begin{cases}
\dot{\theta}_i=p_i\\
\dot{p}_i=-(J+Km^2)(m_x \sin\theta_i-m_y\cos \theta_i)-\lambda\frac{d}{d\theta_i} V_{coup}(\theta_i-\xi_i)\,.
 \end{cases}
\end{equation}

All simulations follow the protocol below:
\begin{enumerate}
 \item during the time interval $0<t<t_0$ the system is decoupled from the
 reservoir ($\lambda=0$) and it evolves deterministically;
 \item the temperature $T$ of the system is computed by averaging the observable $p_i^2$
 over all particles for $0<t<t_0$;
 \item the temperature of the bath is set equal to $T$;
 \item for $t>t_0$ the coupling is switched on ($\lambda>0$) and the total system evolves
 according to the stochastic position Verlet algorithm for Langevin equations
 discussed in Ref.~\cite{melchionna07}.
\end{enumerate}
The process is repeated for several starting specific energy $\varepsilon$ of the
system. 
 \begin{figure}
  \centering
  \includegraphics[width=0.49\textwidth]{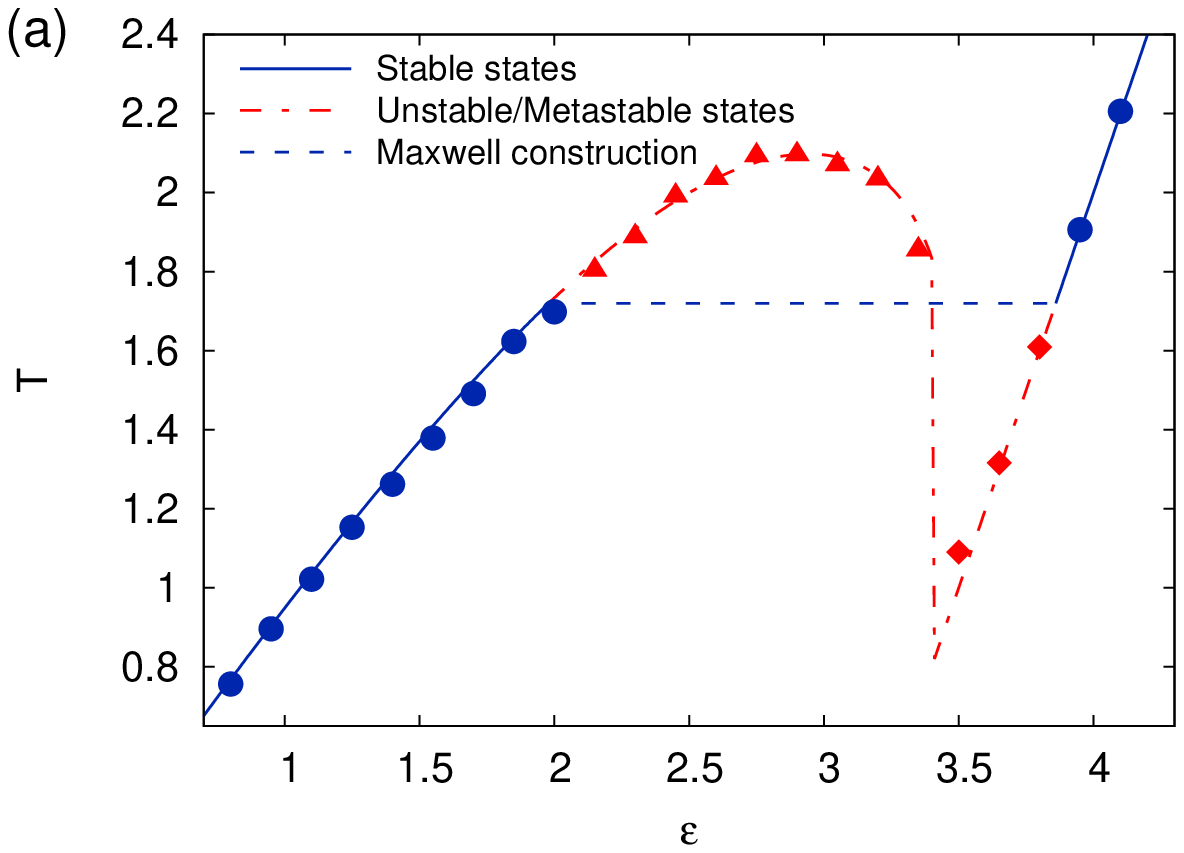}
  \includegraphics[width=0.49\textwidth]{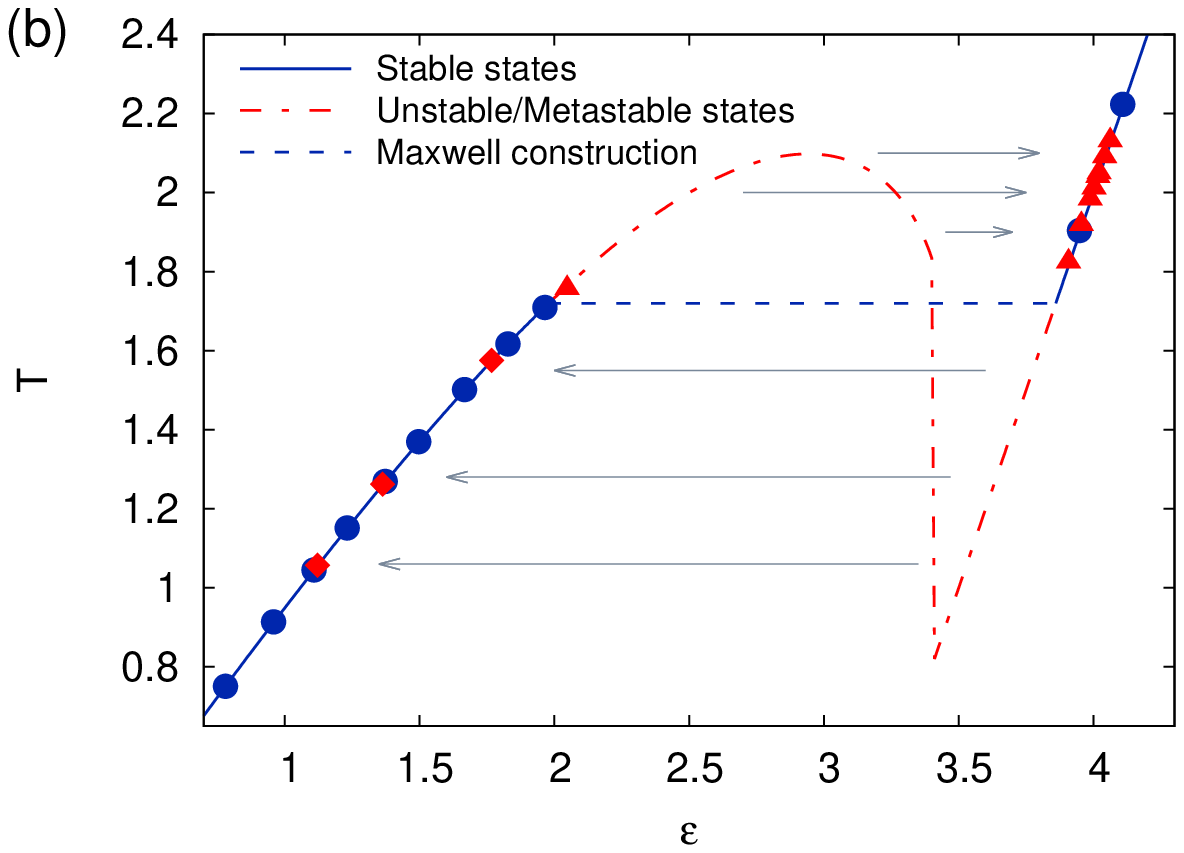}
  \caption{$T(\varepsilon)$ vs $\varepsilon$ for the GHMF model interacting
  with a stochastic thermal bath, before (a) and after (b) the coupling.
  Points have been marked in different ways according to the initial specific
  energy. All simulations have been performed with the following parameters:
  $N=100$, $\lambda=0.02$, $\tau=20$, $t_0=8\cdot10^4$. Averages in the second
  figure have been computed over a time interval $\Delta t=1.6\cdot10^6$, starting
  at $t=1.2\cdot10^6\gg t_0$.}
  \label{fig:bath}
  \end{figure}

The above setting could sound quite unphysical; we remark however
that its study is certainly useful in order to check wether the system can actually reach
the correct equilibrium distribution through the dynamics: such possibility
could be questioned if the system starts from metastable states,
since in this case thermalization times are potentially huge. In addition, this stochastic
approach can give useful insight about the typical waiting times to be expected in the
deterministic simulations.

The results are shown in Fig.~\ref{fig:bath}, in which each point
represents a simulation. 
As long as the system is isolated, its $T(\varepsilon)$ dependence 
is given by the microcanonical caloric curve, which consists not only of stable
states, but also of unstable and metastable ones \cite{campa09}, i.e. states whose $\varepsilon$
does not minimize free energy when $T$ is fixed. This is quite evident in the second
graph of Fig.~\ref{fig:bath}: when the system is coupled to the reservoir, after
some time it reaches the ``true'' equilibrium state at the same temperature
(which is fixed by the bath) but with a different specific energy. For metastable
states this process can take, as it is well known, very long times even for a relatively
small number of particle, and this explains the residual point in the ``forbidden''
branch of the curve.

We stress that this simple stochastic approach clearly shows that, at least for
this particular choice of the physical parameters, dynamics does
select the correct equilibrium distribution (in accessible comuputational times).
This consideration is very important, since it suggests the possibility of similar results
also in deterministic simulations.

 \subsection{Hamiltonian reservoir with short-range interactions}
 The following protocol simulates a thermal bath by using an Hamiltonian system. In
 a more general fashion it has been already introduced in Ref.~\cite{baldovin06} in order to
 study the non equilibrium behavior of the Hamiltonian Mean Field model (system
 \eqref{ham} with $K=0$).
 The reservoir consists of a chain of $N_{res}\gg N$ first-neighbors rotators;
 $N$ of them, randomly chosen, are in turn coupled to the system, trough the
 $\lambda V_{coup}(\delta)$ pair potential (see Eq.~\eqref{vcoup}). Let us remark that in Ref.~\cite{baldovin06}
 each particle of the system was in contact with $S$ particles in the bath;
 choosing $S\propto N^{-1/2}$, one reproduces the ``surface-like'' effect
 in the thermodynamic limit. Here we are considering the case $S=1$,
 with the additional constraint that each rotator of the reservoir can be coupled to
 no more than one particle of the system.
 \\
 The total Hamiltonian is:
 \begin{equation}
  H_{tot}=H_N(\{\theta_i,p_i\})+ \sum_{i=1}^{N_{res}}\frac{\pi_i^2}{2}+\gamma\sum_{i=1}^{N_{res}+1}(1-\cos(\xi_i-\xi_{i-1}))+
  \lambda \sum_{i=1}^{N}V_{coupl}(\xi_{r_i}-\theta_i)
 \end{equation}
 where $\{\xi_i,\pi_i\}$ are the coordinates of the particles in the reservoir 
 ($\xi_0\equiv\xi_{N+1}\equiv 0$) and $\{r_i\}$ are distinct integers randomly
 chosen in the interval $[1,N_{res}]$.
 \begin{figure}
 \centering
 \includegraphics[width=0.49\linewidth]{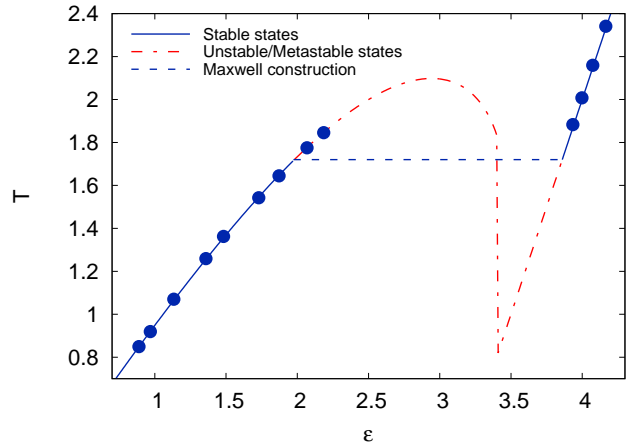}
 \caption{$T(\varepsilon)$ vs $\varepsilon$ for the GHMF model interacting with
 a Hamiltonian reservoir with short-range interactions. Parameters:
 $N=80$, $N_{res}=800$, $\gamma=10$, $\lambda=0.02$. Averages have been computed over a time
 interval $\Delta t=4\cdot10^5$, after $t=8\cdot10^5$ time units.}
 \label{fig:shortshort}
\end{figure}
Simulating the total Hamiltonian at different energies $E_{tot}$, we can sketch the
$T(\varepsilon)$ dependence for the GHMF system. Fig.~\ref{fig:shortshort} shows
that also in this case, once equilibrium has been reached, canonical ensemble
provides the correct statistical description (besides some long-lasting metastable states).
As already noticed in Ref.~\cite{baldovin06}, thermal equilibrium is not ``assumed''
by the simulation protocol (as it happens when stochastic terms are involved),
instead it is reached by the system in a rather physical way.

\subsection{GHMF reservoir}
It could be not completely obvious what does it happen when the reservoir is constituted
by another, larger, GHMF system. In Ref.~\cite{ramirezhernandez09} violations of the zero-th
law of Thermodynamics have been found for a long-range interacting model in which,
as in the GHMF, statistical ensembles are not equivalent; it has been shown that,
if two isolated systems with equal size share the same temperature $T_1$, but their
specific heat is negative, they will reach a different temperature $T_2$ when
coupled each other.

\begin{figure}
 \centering
 \includegraphics[width=0.49\textwidth]{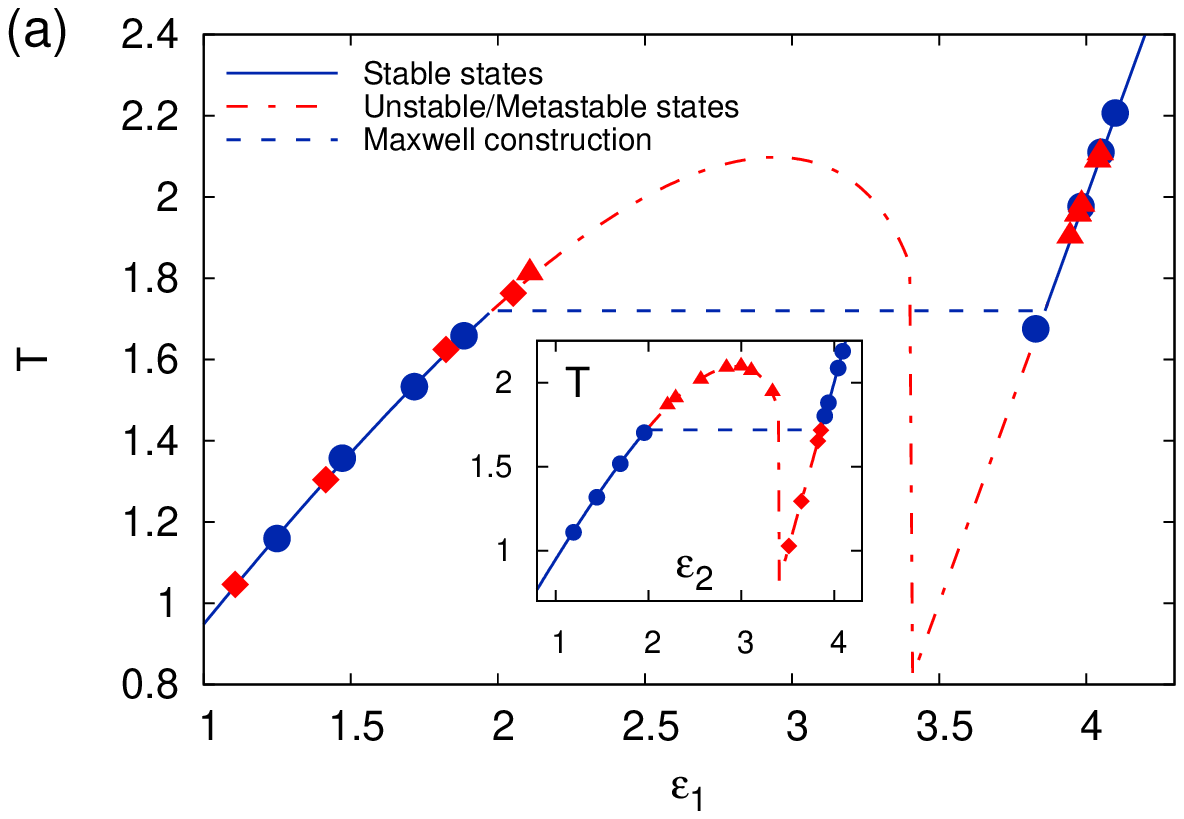}
 \includegraphics[width=0.49\textwidth] {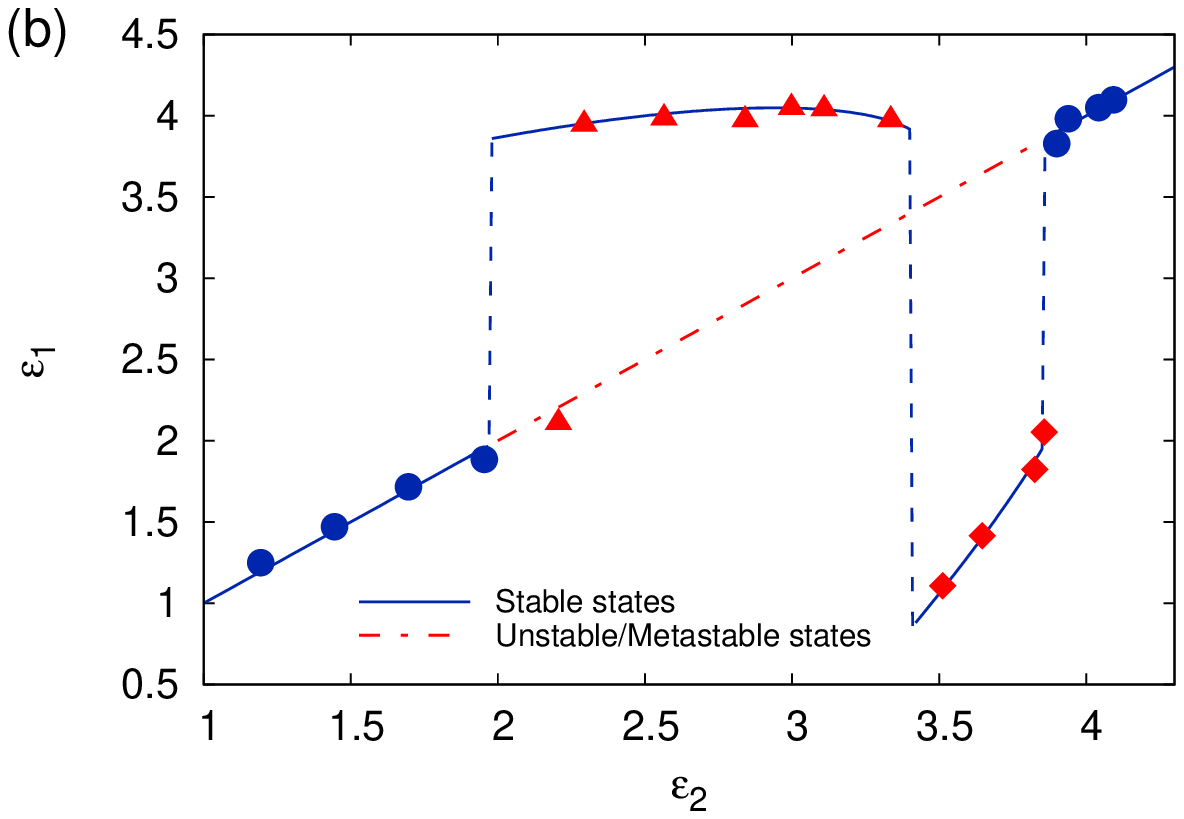}
 \caption{Case of a GHMF system coupled to a GHMF reservoir. (a) Caloric curve for the system, $T(\varepsilon_1)$~vs~$\varepsilon_1$, (main plot) where $T$ is estimated by a direct average $\langle p_i^2 \rangle$ on the particles of the system; the caloric curve for the reservoir, $T(\varepsilon_2)$~vs~$\varepsilon_2$, is also shown (inset). (b) Relation between $\varepsilon_1$ and $\varepsilon_2$. Points are marked in different ways according to the average energy of the reservoir in the considered simulation. Parameters: $N=80$, $N_{res}=800$, $\lambda=0.02$. Averages have been computed over a time interval $\Delta t = 4.0\cdot 10^5$.}
 \label{fig:shortlong}
\end{figure}

It can be easily seen, trough a
microcanonical approach quite similar to the one used in Ref.~\cite{ramirezhernandez09}, that this is
not the case when the ratio $N_2/N_1$ between the sizes of the two systems is very high: in 
such situation the temperature of the larger one does not change significantly,
while, as expected, the thermodynamic behavior of the smaller one is described 
by the canonical ensemble. Indeed, if one defines $\alpha \equiv N_1/(N_1+N_2)$ and
indicates by $\varepsilon_1$ and $\varepsilon_2$ the specific energies of the two
systems, the most probable value of $\varepsilon_1$ at
fixed total energy $E$ can be computed in general by maximizing the total entropy
\begin{equation}
\label{eq:entropy}
s_{tot}(\varepsilon_1,\varepsilon_2)=\alpha s(\varepsilon_1)+(1-\alpha)s(\varepsilon_2)
\end{equation}
with the constraint $\varepsilon_{tot}\equiv E/(N_1+N_2)=\alpha \varepsilon_1+(1-\alpha)\varepsilon_2$,
where $s(\varepsilon)$ is the entropy of the GHMF model. 
Critical points of entropy \eqref{eq:entropy} are obtained for values of $\varepsilon_1$
such that the temperatures of the two subsystems are equal, i.e.:
\begin{equation}
\label{eq:max_ent}
 s'(\varepsilon_1)-s'\left(\frac{\varepsilon_{tot}-\alpha \varepsilon_1}{1-\alpha}\right)=0\,
\end{equation}
Anyway, if different solutions $\varepsilon_1^{(n)}$, $n=1,2,...$ of Eq.~\eqref{eq:max_ent} do exist
(i.e. if $s(\varepsilon)$ is not a strictly concave function), the one that corresponds
to the stable equilibrium, $\varepsilon_1^*$, must fulfill
\begin{equation}
 s_{tot}\left(\varepsilon_1^*,\frac{\varepsilon_{tot}-\alpha \varepsilon_1^*}{1-\alpha}\right)\ge
 s_{tot}\left(\varepsilon_1^{(n)},\frac{\varepsilon_{tot}-\alpha \varepsilon_1^{(n)}}{1-\alpha}\right)\; \forall n\,.
\end{equation}
The above inequality can be studied in the $\alpha \ll 1$ limit with a first-order
expansion. One gets
\begin{equation}
 \alpha\left[s(\varepsilon_1^*)- s(\varepsilon_1^{(n)})\right]\ge
 s'(\varepsilon_{tot})\alpha (\varepsilon_{tot}-\varepsilon^{(n)})- 
 s'(\varepsilon_{tot})\alpha(\varepsilon_{tot}-\varepsilon_1^*)
\end{equation}
which immediately leads to the integral condition
\begin{equation}
 \int_{\varepsilon_1^*}^{\varepsilon_1^{(n)}}T^{-1}(\varepsilon')d \varepsilon' 
 \le T^{-1}(\varepsilon_{tot})(\varepsilon_1^{(n)}-\varepsilon_1^*)\; \forall n\,
\end{equation}
because of the relation $T^{-1}(\varepsilon)\equiv s'(\varepsilon)$.
The above condition is nothing but the Maxwell construction; one can therefore
conclude that in the limit $\alpha \ll 1$, i.e. when it is possible to identify a
reservoir composed of $N_{res}=N_2$ particles and a small system made of $N=N_1$ rotators coupled to it,
with $N\ll N_{res}$, the equilibrium behavior of the second is described by the canonical ensemble at temperature $T(\varepsilon_{tot})$.
Since
\begin{equation}
 \varepsilon_2=\frac{\varepsilon_{tot}-\alpha \varepsilon_1}{1-\alpha}\approx  
 \varepsilon_{tot}+\alpha (\varepsilon_{tot}- \varepsilon_1) +O(\alpha^2)
\end{equation}
it is also proved that $T(\varepsilon_{tot})\approx T(\varepsilon_{2}) + O(\alpha)$
(if $\varepsilon_{tot}$ is not too close to a microcanonical phase transition),
i.e. the temperature of the small system is determined by the one of the reservoir,
as expected, even if the reservoir is in an ``unstable'' state with negative specific heat.
\\
The above considerations can be tested by numerical simulations on a system of the kind:
\begin{equation}
 H_{tot}=H_{N}(\{\theta_i, p_i\}) + H_{N_{res}}(\{\xi_i, \pi_i\})+ \lambda \sum_{i=1}^{N}V_{coup}(\theta_i-\xi_i)
\end{equation}
where $H_N$ and $V_{coup}$ have been defined in Eq.~\eqref{ham}~and~\eqref{vcoup}. In Fig.~\ref{fig:shortlong}, panel (a),
we see that the $T(\varepsilon)$ dependence for the small system is in a rather good
agreement with the theoretical prediction,  where $T$ is estimated by the average $\langle p_i^2 \rangle$ on the $N$ particles of the small system itself. Not surprisingly, in some cases the system is
trapped in a metastable state.  As expected, the reservoir (inset) can assume every 
specific energy at equilibrium, even those leading to negative specific heat. Panel (b) of Fig.~\ref{fig:shortlong} shows the relation between the specific energies of the bath ($\varepsilon_2$) and that of the system  ($\varepsilon_1$), compared to the theoretical curve. 

\newpage
\section{Equilibrium behavior of a weakly interacting portion of a mean-field system}
\label{sec:weak}
Let us consider the (quite reasonable) situation in which a mean-field interacting system
is split into two parts $\mathcal{S}_1$ and $\mathcal{S}_2$, in such a way
that the effective mean field acting on each particle of $\mathcal{S}_1$ depends
strongly on the degrees of freedom $\mathbf{X}_1$ of $\mathcal{S}_1$ itself 
and weakly on those of $\mathcal{S}_2$ (i.e. $\mathbf{X}_2$), and vice versa.
In real physical systems this could be obtained
by some kind of screening between the two parts, or by simply distancing them to
a range in which mean field interactions are no more a valid approximation.

Consider the case of GHMF model, and call $\mathbf{m}_1(\mathbf{X}_1)$
and $\mathbf{m}_2(\mathbf{X}_2)$ the magnetization vectors of the two subsystems,
whose components are
defined according to Eq.~\eqref{magn}. It is reasonable to assume that the
effective fields acting on $\mathcal{S}_1$ and $\mathcal{S}_2$ can be described by
\begin{equation}
\begin{cases}
  \mathbf{m}_1^*(\mathbf{X}_1,\mathbf{X}_2)=(1-\lambda)\mathbf{m}_1(\mathbf{X}_1)+\lambda \mathbf{m}(\mathbf{X}_1,\mathbf{X}_2)\\
 \mathbf{m}_2^*(\mathbf{X}_1,\mathbf{X}_2)=(1-\lambda)\mathbf{m}_2(\mathbf{X}_2)+\lambda \mathbf{m}(\mathbf{X}_1,\mathbf{X}_2)
\end{cases}
\end{equation}
where $\mathbf{m}(\mathbf{X}_1,\mathbf{X}_2)$ is the mean field of the total system without splitting
and $\lambda\in [0,1]$ is a real parameter which quantifies the interaction, so
that $\lambda=0$ when the subsystems are completely isolated and $\lambda=1$
when there's no screening at all.
The total Hamiltonian reads
\begin{equation}\label{ham_eff}
 H_{tot}(\mathbf{X}_1,\mathbf{X}_2)=\sum_{i=1}^{N_1+N_2}\frac{p_i^2}{2} + N_1u(m_1^*)+ N_2u(m_2^*)
\end{equation}
where $N_1$ and $N_2$ are the number of particles in $\mathcal{S}_1$ and $\mathcal{S}_2$, 
and 
$$
u(x)\equiv \frac{J}{2}(1-x^2)+\frac{K}{4}(1-x^4)\,.
$$
The equilibrium properties for given values of $\lambda$ and $\alpha\equiv N_1/(N_1+N_2)$
of this Hamiltonian system can be derived exactly, in the thermodynamic
limit, by using large deviations techniques (see Appendix).

Let us note that, since long-range interactions are involved, one can introduce
different definitions for the energy of $\mathcal{S}_1$, depending on which extent the
non-negligible interactions with $\mathcal{S}_2$ are taken into account. In this
context, anyway, the following microcanonical average 
\begin{equation}\label{energy}
E_1=\langle\sum_{i=1}^{N_1}\frac{p_i^2}{2} + N_1u(m_1^*) \rangle
\end{equation}
seems to be a quite reasonable choice.

Let us focus first on the $\lambda \ll 1$ case. Eq.~\eqref{ham_eff} can be rewritten
in the form 
\begin{equation}\label{ham_split}
 H_{tot}(\mathbf{X}_1,\mathbf{X}_2)=H_{N_1}(\mathbf{X}_1)+H_{N_2}(\mathbf{X}_1)+ \lambda H_{int}(\mathbf{X}_1,\mathbf{X}_2)
\end{equation}
where $H_N$ is the GHMF model Hamiltonian \eqref{ham} for a system of $N$ particles
and the $\lambda H_{int}$, whose average is negligible with respect to those of $H_{N_1}$
and $H_{N_2}$, includes all interactions terms between the two systems. All interactions
in this system are long-range; nonetheless, in this particular limit, we recover
the conditions that are needed in the well-known derivation of canonical ensemble
from a microcanonical description. Assuming that ergodicity holds, in this
limit one expects a thermodynamic behavior quite similar to those that have
been discussed in Section \ref{sec:baths}.

In the opposite limit, namely $\lambda \lesssim 1$, the energy
range in which $m^*_1\ne m^*_2$ gradually shrinks. Above certain critical value
$\bar{\lambda}(\alpha)$, condition $m^*_1= m^*_2$ (or, equivalently, $m_1=m_2$)
always holds at equilibrium, and for $\lambda=1$, the $T(\varepsilon_1)$ curve will coincide
with the microcanonical one, as long as the above definition of energy is considered.

\begin{figure}
 \centering
       \includegraphics[width=0.49\textwidth]{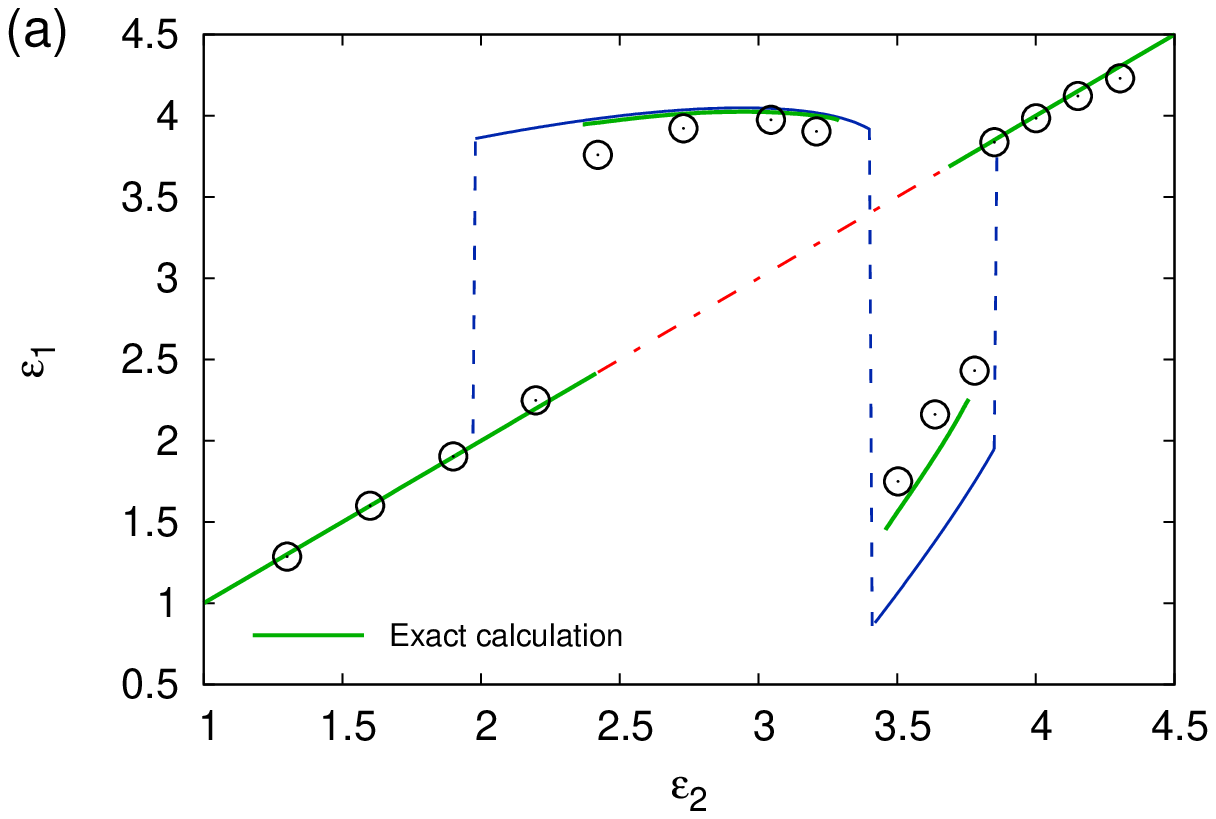}
       \includegraphics[width=0.49\textwidth]{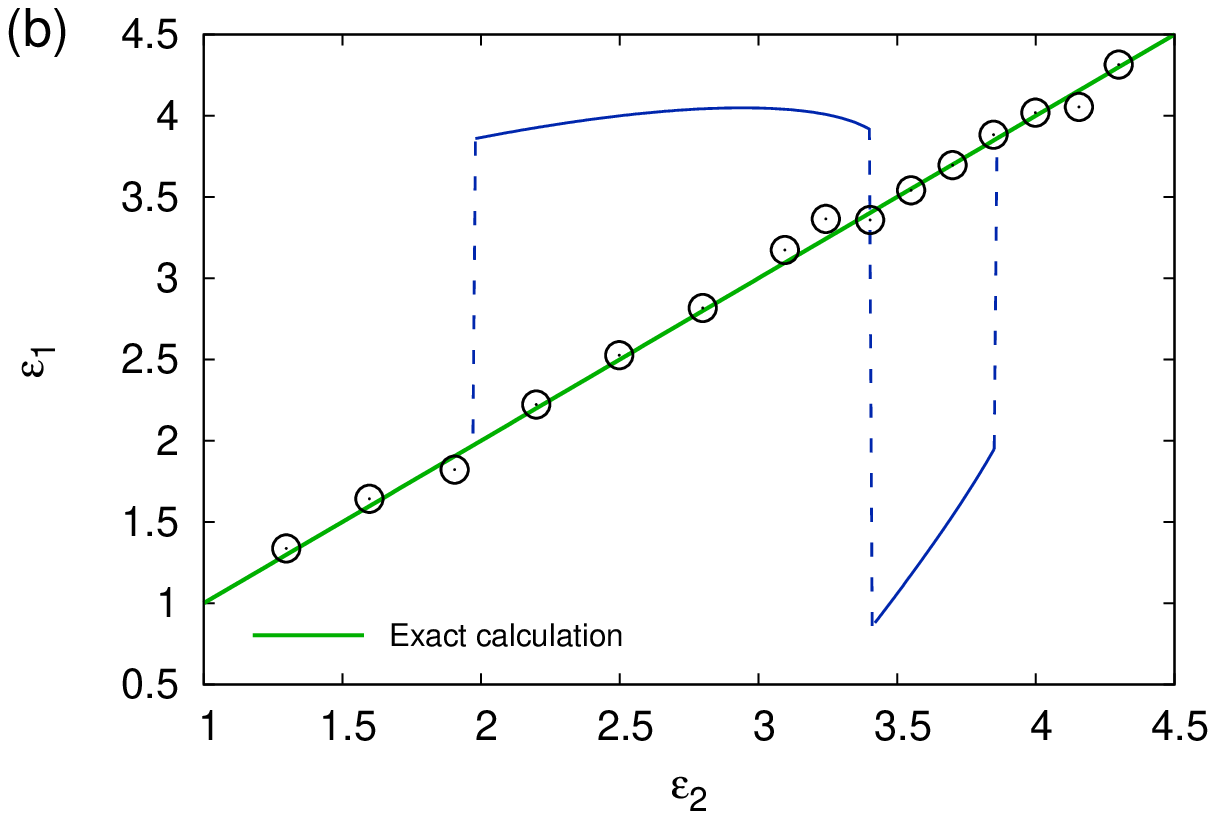}
     \caption{Specific energy $\varepsilon_1$ of the small portion vs specific
     energy $\varepsilon_2$ of the large one, for $\lambda=0.05$ (a) and $\lambda=0.1$ (b). Green solid lines stand for the exact solutions computed as described in the Appendix.
     Equilibrium behavior of the system in the
     microcanonical (red) and canonical (blue) ensembles are also shown for comparison.
     Parameters: $N_1=50$, $N_2=800$ ($\alpha=1/17$), $\lambda=0.02$. Averages
     have been computed over time intervals $\Delta t=3.5\cdot 10^7$ (left) and 
     $\Delta t=10^7$ (right).}
 \label{fig:longlong}
\end{figure}

In Fig.~\ref{fig:longlong} the two situations are shown for a fixed (small) value of
$\alpha$, and numerical simulations are compared to analytical calculations.
\\
As far as mean field interactions are concerned, some general considerations about the
thermodynamics of a small piece $\mathcal{S}_1$ of the total system $\mathcal{S}_{tot}$
can be outlined. If definition \eqref{energy} is considered, the caloric curve of
$\mathcal{S}_1$ is the same of an isolated system: in particular negative specific
heat can be observed, because of the action of the mean field of $\mathcal{S}_{tot}$
which keeps the subsystem in unstable energy regions; when the effect of the total
mean field is weakened trough some screening, but not enough to prevent heat
exchange between the two subsystems, unstable and metastable states are no more
accessible for $\mathcal{S}_1$ and the canonical description is recovered in the
limit.
\\
On the other hand, if one defines the energy of the subsystem as the sum of all 
terms of the total Hamiltonian which depend on $\mathbf{X}_1$ only, the caloric
curve tends to the one of the ideal gas, since average kinetic energy is of order
$\alpha N$ and the average potential energy is of order $\alpha^2 N$, because of Kac's 
prescription \cite{campa09}.
\newpage
\section{Final remarks}
\label{sec:conclusions}
In this paper we have investigated several different physical situations in which the equilibrium
behavior of a long-range interacting system with ensemble inequivalence
is described by the canonical distribution. The aim of such approach is to clarify the
physical interpretation of this statistical ensemble, which can always be defined
from a mathematical point of view.

First we have studied, by numerical simulations,
the case in which a small system is in contact with a large reservoir; then
we have analyzed the equilibrium behavior of a small portion of a mean-field system,
partially isolated from the rest of it by some kind of screening.
In both cases, the studied degrees of freedom interact weakly with the remaining
part of the system; nonetheless, energy can still be exchanged,
so that the larger part of the system determines the temperature of the smaller one.
This is indeed a physically relevant way to construct the canonical ensemble.

Our results show that the canonical distribution is physically meaningful also
when inequivalence of statistical ensemble is present, as far as the 
above conditions hold. Since such assumptions are verified 
in rather interesting cases, the usage of canonical ensemble 
for long-range interacting systems seems quite natural and fully justified
from a physical point of view.

\section{Appendix}

In this Appendix we use large deviations techniques to investigate the 
equilibrium behavior, in the thermodynamic limit, of the Hamiltonian system
\eqref{ham_eff}. Large deviations are a well-known tool for the study of mean-field
systems \cite{patelli_vulpiani14}. This approach can be used if the Hamiltonian depends only
on $n\ll N$ mean quantities $\mu_j(\mathbf{X})$ with the form $\mu_j=\sum_{i=1}^Ng(q_i, p_i)$,
$j=1,...,n$, or if the energy contribution of other terms is negligible in the thermodynamic
limit. With the above assumptions it is possible to
compute the so-called entropy of macrostates
\begin{equation}\label{entropy_macro}
 \bar{s}(\bar{\mu}_1,...,\bar{\mu}_n)\equiv \frac{1}{N}\ln \int d\mathbf{X}\,
\delta(\mu_1(\mathbf{X})-\bar{\mu}_1)\delta(\mu_2(\mathbf{X})-\bar{\mu}_2)... 
\delta(\mu_n(\mathbf{X})-\bar{\mu}_n)\,,
\end{equation}
which is maximal in the equilibrium macrostates of the microcanonical ensemble.
Even if Hamiltonian \eqref{ham_eff} is not in the requested form, its $\bar{s}(\bar{\mu}_1,...,\bar{\mu}_n)$
can be easily computed.
Indeed, Hamiltonian \eqref{ham_eff} can be written as
\begin{equation}
H_{tot}(\mathbf{X}_1,\mathbf{X}_2)=\bar{H}(\kappa_1, m_{1x}, m_{1y},\kappa_2, m_{2x}, m_{2y})=\bar{H}(\mathbf{w}_1,\mathbf{w}_2)
\end{equation}
where
$$
\begin{aligned}
\kappa_{1}&=\frac{1}{N_1}\sum_{i=1}^{N_1}p_i^2	&	m_{1x}&=\frac{1}{N_1}\sum_{i=1}^{N_1}\cos\theta_i	&	m_{1y}&=\frac{1}{N_1}\sum_{i=1}^{N_1}\sin\theta_i\\
\kappa_{2}&=\frac{1}{N_2}\sum_{i=N_1+1}^{N_1+N_2}p_i^2	&	m_{2x}&=\frac{1}{N_2}\sum_{i=N_1+1}^{N_1+N_2}\cos\theta_i	&	m_{2y}&=\frac{1}{N_2}\sum_{i=N_1+1}^{N_1+N_2}\sin\theta_i
\end{aligned}
$$
and
$$
\mathbf{w}_1=(\kappa_{1}, m_{1x}, m_{1y})\quad\quad \mathbf{w}_2=(\kappa_{2}, m_{2x}, m_{2y})
$$
once one recognizes that $\mathbf{m}=\alpha\mathbf{m}_1+(1-\alpha)\mathbf{m}_2$,
where $\alpha=\frac{N_1}{N_1+N_2}$.
The microcanonical entropy, depending on total energy $E$,
can be written as
\begin{equation}
 \begin{aligned}
  S_{tot}(E)&=\ln\int d\mathbf{X}_1 d\mathbf{X}_2\, \delta(H(\mathbf{X}_1,\mathbf{X}_2)-E)\\
  &= \ln\int d\bar{\mathbf{w}}_{1} d\bar{\mathbf{w}}_{2} \,d\mathbf{X}_1 d\mathbf{X}_2\, 
  \delta(\bar{H}(\bar{\mathbf{w}}_{1}, \bar{\mathbf{w}}_{2})-E)\,\delta(\bar{\mathbf{w}}_{1}-\mathbf{w}_{1}(\mathbf{X}_1))\delta(\bar{\mathbf{w}}_{2}-\mathbf{w}_{2}(\mathbf{X}_2))\\
  &=\ln\int d\bar{\mathbf{w}}_{1} d\bar{\mathbf{w}}_{2} \,\delta(\bar{H}(\bar{\mathbf{w}}_{1}, \bar{\mathbf{w}}_{2})-E)\,
  \exp\left[N \bar{s}(\bar{\mathbf{w}}_{1}, \bar{\mathbf{w}}_{2}) \right]
 \end{aligned}
\end{equation}
with
\begin{equation}\label{entro_macro}
\bar{s}(\bar{\mathbf{w}}_{1}, \bar{\mathbf{w}}_{2})\equiv\frac{1}{N}\ln\int d\mathbf{X}_1 d\mathbf{X}_2\, 
\delta(\bar{\mathbf{w}}_{1}-\mathbf{w}_{1}(\mathbf{X}_1))\delta(\bar{\mathbf{w}}_{2}-\mathbf{w}_{2}(\mathbf{X}_2))\,.
\end{equation}
Since
$$
\frac{S_{tot}(E)}{N}\approx  \sup_{(\bar{\mathbf{w}}_{1}, \bar{\mathbf{w}}_{2})| \bar{H}(\bar{\mathbf{w}}_{1}, \bar{\mathbf{w}}_{2})=E}\bar{s}(\bar{\mathbf{w}}_{1}, \bar{\mathbf{w}}_{2})
$$
assuming that one can compute the entropy of macrostate \eqref{entro_macro}, the
problem of computing the microcanonical entropy is thus reduced to that of finding
a constrained supremum.
This is indeed the case, since
\begin{equation}
\begin{aligned}
\bar{s}(\bar{\mathbf{w}}_{1}, \bar{\mathbf{w}}_{2})&=\frac{\alpha}{N_1}\ln\int d\mathbf{X}_1\,
\delta(\bar{\mathbf{w}}_{1}-\mathbf{w}_{1}(\mathbf{X}_1))+\frac{1-\alpha}{N_2}\ln\int d\mathbf{X}_2\,
\delta(\bar{\mathbf{w}}_{2}-\mathbf{w}_{2}(\mathbf{X}_2))\\
&\equiv\alpha \tilde{s}_1(\bar{\kappa}_{1},\bar{m}_{1x},\bar{m}_{1y})+
(1-\alpha)\tilde{s}_2(\bar{\kappa}_{2},\bar{m}_{2x},\bar{m}_{2y})
\end{aligned}
\end{equation}
where $\tilde{s}(\bar{\kappa},\bar{m}_{x},\bar{m}_{y})$ is the entropy of macrostates
for the GHMF model, that can be computed as discussed in Ref.~\cite{campa09}.
\\
The final result is
\begin{equation}\label{s_final}
\begin{aligned}
 \bar{s}(\kappa_1, \kappa_2, m_1, m_2)&=\frac{1}{2}(1+\ln \pi)+ \frac{\alpha}{2}\ln(2\kappa_1)+ \frac{1-\alpha}{2}\ln(2\kappa_2)\\
 &+ \alpha \left[ -m_1 B_{inv}({m_1})+\ln[I_0(B_{inv}({m_1}))]\right]\\
 &+(1-\alpha) \left[ -m_2 B_{inv}({m_2})+\ln[I_0(B_{inv}({m_2}))]\right]
 \end{aligned}
\end{equation}
where $I_n(x)$ is the $n$-th modified Bessel function of the first kind and $B_{inv}(x)$
is the inverse of $B(x)\equiv I_1(x)/I_0(x)$. Let us notice that, due to the form of the
Hamiltonian, in entropy \eqref{s_final} only the moduli $m_1$ and $m_2$ of
vectors $\mathbf{m}_1$, $\mathbf{m}_2$ appear: the task of maximizing this quantity
with the constraint $\bar{H}(\bar{\kappa}_1,...,\bar{m}_{2y})=E$
can be performed numerically.

\acknowledgments{I would like to thank A. Campa, L. Cerino and A. Vulpiani for helpful discussions and useful comments on the manuscript.}

\bibliography{biblio}

\begin{thebibliography}{27}
\expandafter\ifx\csname natexlab\endcsname\relax\def\natexlab#1{#1}\fi
\expandafter\ifx\csname bibnamefont\endcsname\relax
  \def\bibnamefont#1{#1}\fi
\expandafter\ifx\csname bibfnamefont\endcsname\relax
  \def\bibfnamefont#1{#1}\fi
\expandafter\ifx\csname citenamefont\endcsname\relax
  \def\citenamefont#1{#1}\fi
\expandafter\ifx\csname url\endcsname\relax
  \def\url#1{\texttt{#1}}\fi
\expandafter\ifx\csname urlprefix\endcsname\relax\def\urlprefix{URL }\fi
\providecommand{\bibinfo}[2]{#2}
\providecommand{\eprint}[2][]{\url{#2}}

\bibitem[{\citenamefont{Huang}(1988)}]{huang1988}
\bibinfo{author}{\bibfnamefont{K.}~\bibnamefont{Huang}},
  \emph{\bibinfo{title}{{Statistical Mechanics}}} (\bibinfo{publisher}{John
  Wiley \& Sons}, \bibinfo{address}{New York}, \bibinfo{year}{1988}).

\bibitem[{\citenamefont{Touchette}(2009)}]{touchette09}
\bibinfo{author}{\bibfnamefont{H.}~\bibnamefont{Touchette}},
  \bibinfo{journal}{Phys. Rep.} \textbf{\bibinfo{volume}{478}},
  \bibinfo{pages}{1 } (\bibinfo{year}{2009}).

\bibitem[{\citenamefont{Ma}(1985)}]{ma85}
\bibinfo{author}{\bibfnamefont{S.-K.} \bibnamefont{Ma}},
  \emph{\bibinfo{title}{Statistical Mechanics}} (\bibinfo{publisher}{World
  Scientific}, \bibinfo{address}{Phyladelphia Singapore},
  \bibinfo{year}{1985}).

\bibitem[{\citenamefont{Sethna}(2005)}]{sethna05}
\bibinfo{author}{\bibfnamefont{J.~P.} \bibnamefont{Sethna}},
  \emph{\bibinfo{title}{Statistical Mechanics: Entropy, Order Parameters and
  Complexity}} (\bibinfo{publisher}{Oxford University Press},
  \bibinfo{address}{Oxford}, \bibinfo{year}{2005}).

\bibitem[{\citenamefont{Seifert}(2012)}]{seifert12}
\bibinfo{author}{\bibfnamefont{U.}~\bibnamefont{Seifert}},
  \bibinfo{journal}{Rep. Prog. Phys.} \textbf{\bibinfo{volume}{75}},
  \bibinfo{pages}{126001} (\bibinfo{year}{2012}).

\bibitem[{\citenamefont{Cerino et~al.}(2015)\citenamefont{Cerino, Puglisi, and
  Vulpiani}}]{cerino15}
\bibinfo{author}{\bibfnamefont{L.}~\bibnamefont{Cerino}},
  \bibinfo{author}{\bibfnamefont{A.}~\bibnamefont{Puglisi}}, \bibnamefont{and}
  \bibinfo{author}{\bibfnamefont{A.}~\bibnamefont{Vulpiani}},
  \bibinfo{journal}{Phys. Rev. E} \textbf{\bibinfo{volume}{91}},
  \bibinfo{pages}{P12002} (\bibinfo{year}{2015}).

\bibitem[{\citenamefont{Puglisi et~al.}(2017)\citenamefont{Puglisi, Sarracino,
  and Vulpiani}}]{puglisi17}
\bibinfo{author}{\bibfnamefont{A.}~\bibnamefont{Puglisi}},
  \bibinfo{author}{\bibfnamefont{A.}~\bibnamefont{Sarracino}},
  \bibnamefont{and} \bibinfo{author}{\bibfnamefont{A.}~\bibnamefont{Vulpiani}},
  \bibinfo{journal}{Phys. Rep.} \textbf{\bibinfo{volume}{709-710}},
  \bibinfo{pages}{1} (\bibinfo{year}{2017}).

\bibitem[{\citenamefont{Dauxois et~al.}(2002)\citenamefont{Dauxois, Ruffo,
  Arimondo, and Wilkens}}]{dauxois02}
\bibinfo{editor}{\bibfnamefont{T.}~\bibnamefont{Dauxois}},
  \bibinfo{editor}{\bibfnamefont{S.}~\bibnamefont{Ruffo}},
  \bibinfo{editor}{\bibfnamefont{E.}~\bibnamefont{Arimondo}}, \bibnamefont{and}
  \bibinfo{editor}{\bibfnamefont{M.}~\bibnamefont{Wilkens}}, eds.,
  \emph{\bibinfo{title}{Dynamics and Thermodynamics of Systems with Long Range
  Interactions}} (\bibinfo{publisher}{Springer-Verlag},
  \bibinfo{address}{Berlin Heidelberg}, \bibinfo{year}{2002}).

\bibitem[{\citenamefont{Campa et~al.}(2009)\citenamefont{Campa, Dauxois, and
  Ruffo}}]{campa09}
\bibinfo{author}{\bibfnamefont{A.}~\bibnamefont{Campa}},
  \bibinfo{author}{\bibfnamefont{T.}~\bibnamefont{Dauxois}}, \bibnamefont{and}
  \bibinfo{author}{\bibfnamefont{S.}~\bibnamefont{Ruffo}},
  \bibinfo{journal}{Phys. Rep.} \textbf{\bibinfo{volume}{480}},
  \bibinfo{pages}{57 } (\bibinfo{year}{2009}).

\bibitem[{\citenamefont{Chavanis}(2006)}]{chavanis06}
\bibinfo{author}{\bibfnamefont{P.~H.} \bibnamefont{Chavanis}},
  \bibinfo{journal}{International Journal of Modern Physics B}
  \textbf{\bibinfo{volume}{20}}, \bibinfo{pages}{3113} (\bibinfo{year}{2006}).

\bibitem[{\citenamefont{Sire and Chavanis}(2004)}]{sire04}
\bibinfo{author}{\bibfnamefont{C.}~\bibnamefont{Sire}} \bibnamefont{and}
  \bibinfo{author}{\bibfnamefont{P.-H.} \bibnamefont{Chavanis}},
  \bibinfo{journal}{Phys. Rev. E} \textbf{\bibinfo{volume}{69}},
  \bibinfo{pages}{066109} (\bibinfo{year}{2004}).

\bibitem[{\citenamefont{Antoni and Ruffo}(1995)}]{antoni95}
\bibinfo{author}{\bibfnamefont{M.}~\bibnamefont{Antoni}} \bibnamefont{and}
  \bibinfo{author}{\bibfnamefont{S.}~\bibnamefont{Ruffo}},
  \textbf{\bibinfo{volume}{52}}, \bibinfo{pages}{2361} (\bibinfo{year}{1995}).

\bibitem[{\citenamefont{Chavanis}(2014)}]{chavanis14}
\bibinfo{author}{\bibfnamefont{P.-H.} \bibnamefont{Chavanis}},
  \bibinfo{journal}{The European Physical Journal B}
  \textbf{\bibinfo{volume}{87}}, \bibinfo{pages}{120} (\bibinfo{year}{2014}).

\bibitem[{\citenamefont{{Gross}}(2002)}]{gross02}
\bibinfo{author}{\bibfnamefont{D.~H.~E.} \bibnamefont{{Gross}}},
  \bibinfo{journal}{Phys. Chem. Chem. Phys.} \textbf{\bibinfo{volume}{4}},
  \bibinfo{pages}{863} (\bibinfo{year}{2002}).

\bibitem[{\citenamefont{Gross}(2002)}]{gross_campa02}
\bibinfo{author}{\bibfnamefont{D.~H.~E.} \bibnamefont{Gross}}, in
  \emph{\bibinfo{booktitle}{Dynamics and Thermodynamics of Systems with Long
  Range Interactions}}, edited by
  \bibinfo{editor}{\bibfnamefont{T.}~\bibnamefont{Dauxois}},
  \bibinfo{editor}{\bibfnamefont{S.}~\bibnamefont{Ruffo}},
  \bibinfo{editor}{\bibfnamefont{E.}~\bibnamefont{Arimondo}}, \bibnamefont{and}
  \bibinfo{editor}{\bibfnamefont{M.}~\bibnamefont{Wilkens}}
  (\bibinfo{publisher}{Springer-Verlag}, \bibinfo{address}{Berlin Heidelberg},
  \bibinfo{year}{2002}), pp. \bibinfo{pages}{23--44}.

\bibitem[{\citenamefont{Padmanabhan}(1990)}]{padmanabhan90}
\bibinfo{author}{\bibfnamefont{T.}~\bibnamefont{Padmanabhan}},
  \bibinfo{journal}{Physics Reports} \textbf{\bibinfo{volume}{188}},
  \bibinfo{pages}{285} (\bibinfo{year}{1990}).

\bibitem[{\citenamefont{Chavanis}(2002)}]{chavanis_campa02}
\bibinfo{author}{\bibfnamefont{H.~P.} \bibnamefont{Chavanis}}, in
  \emph{\bibinfo{booktitle}{Dynamics and Thermodynamics of Systems with Long
  Range Interactions}}, edited by
  \bibinfo{editor}{\bibfnamefont{T.}~\bibnamefont{Dauxois}},
  \bibinfo{editor}{\bibfnamefont{S.}~\bibnamefont{Ruffo}},
  \bibinfo{editor}{\bibfnamefont{E.}~\bibnamefont{Arimondo}}, \bibnamefont{and}
  \bibinfo{editor}{\bibfnamefont{M.}~\bibnamefont{Wilkens}}
  (\bibinfo{publisher}{Springer-Verlag}, \bibinfo{address}{Berlin Heidelberg},
  \bibinfo{year}{2002}), pp. \bibinfo{pages}{208--289}.

\bibitem[{\citenamefont{Baldovin and Orlandini}(2006)}]{baldovin06}
\bibinfo{author}{\bibfnamefont{F.}~\bibnamefont{Baldovin}} \bibnamefont{and}
  \bibinfo{author}{\bibfnamefont{E.}~\bibnamefont{Orlandini}},
  \textbf{\bibinfo{volume}{96}}, \bibinfo{pages}{240602}
  (\bibinfo{year}{2006}).

\bibitem[{\citenamefont{Baldovin and Orlandini}(2007)}]{baldovin07}
\bibinfo{author}{\bibfnamefont{F.}~\bibnamefont{Baldovin}} \bibnamefont{and}
  \bibinfo{author}{\bibfnamefont{E.}~\bibnamefont{Orlandini}},
  \bibinfo{journal}{Int. J. Mod. Phys. B} \textbf{\bibinfo{volume}{21}},
  \bibinfo{pages}{4000} (\bibinfo{year}{2007}).

\bibitem[{\citenamefont{Baldovin et~al.}(2009)\citenamefont{Baldovin, Chavanis,
  and Orlandini}}]{baldovin09}
\bibinfo{author}{\bibfnamefont{F.}~\bibnamefont{Baldovin}},
  \bibinfo{author}{\bibfnamefont{P.-H.} \bibnamefont{Chavanis}},
  \bibnamefont{and}
  \bibinfo{author}{\bibfnamefont{E.}~\bibnamefont{Orlandini}},
  \bibinfo{journal}{Phys. Rev. E} \textbf{\bibinfo{volume}{79}},
  \bibinfo{pages}{011102} (\bibinfo{year}{2009}).

\bibitem[{\citenamefont{de~Buyl et~al.}(2005)\citenamefont{de~Buyl, Mukamel,
  and Ruffo}}]{debuyl05}
\bibinfo{author}{\bibfnamefont{P.}~\bibnamefont{de~Buyl}},
  \bibinfo{author}{\bibfnamefont{D.}~\bibnamefont{Mukamel}}, \bibnamefont{and}
  \bibinfo{author}{\bibfnamefont{S.}~\bibnamefont{Ruffo}},
  \bibinfo{journal}{AIP Conf. Proc.} \textbf{\bibinfo{volume}{800}},
  \bibinfo{pages}{533} (\bibinfo{year}{2005}).

\bibitem[{\citenamefont{Patelli and Ruffo}(2014)}]{patelli_vulpiani14}
\bibinfo{author}{\bibfnamefont{A.}~\bibnamefont{Patelli}} \bibnamefont{and}
  \bibinfo{author}{\bibfnamefont{S.}~\bibnamefont{Ruffo}}, in
  \emph{\bibinfo{booktitle}{Large Deviations in Physics}}, edited by
  \bibinfo{editor}{\bibfnamefont{A.}~\bibnamefont{Vulpiani}},
  \bibinfo{editor}{\bibfnamefont{F.}~\bibnamefont{Cecconi}},
  \bibinfo{editor}{\bibfnamefont{M.}~\bibnamefont{Cencini}},
  \bibinfo{editor}{\bibfnamefont{A.}~\bibnamefont{Puglisi}}, \bibnamefont{and}
  \bibinfo{editor}{\bibfnamefont{D.}~\bibnamefont{Vergni}}
  (\bibinfo{publisher}{Springer-Verlag}, \bibinfo{address}{Berlin Heidelberg},
  \bibinfo{year}{2014}), pp. \bibinfo{pages}{193--220}.

\bibitem[{\citenamefont{Yamaguchi et~al.}(2004)\citenamefont{Yamaguchi,
  Barr{\'e}, Bouchet, Dauxois, and Ruffo}}]{yamaguchi04}
\bibinfo{author}{\bibfnamefont{Y.~Y.} \bibnamefont{Yamaguchi}},
  \bibinfo{author}{\bibfnamefont{J.}~\bibnamefont{Barr{\'e}}},
  \bibinfo{author}{\bibfnamefont{F.}~\bibnamefont{Bouchet}},
  \bibinfo{author}{\bibfnamefont{T.}~\bibnamefont{Dauxois}}, \bibnamefont{and}
  \bibinfo{author}{\bibfnamefont{S.}~\bibnamefont{Ruffo}},
  \bibinfo{journal}{{Physica A}} \textbf{\bibinfo{volume}{337}},
  \bibinfo{pages}{36} (\bibinfo{year}{2004}).

\bibitem[{\citenamefont{de~Buyl et~al.}(2011)\citenamefont{de~Buyl, Mukamel,
  and Ruffo}}]{debuyl11}
\bibinfo{author}{\bibfnamefont{P.}~\bibnamefont{de~Buyl}},
  \bibinfo{author}{\bibfnamefont{D.}~\bibnamefont{Mukamel}}, \bibnamefont{and}
  \bibinfo{author}{\bibfnamefont{S.}~\bibnamefont{Ruffo}},
  \bibinfo{journal}{Phys. Rev. E} \textbf{\bibinfo{volume}{84}},
  \bibinfo{pages}{061151} (\bibinfo{year}{2011}).

\bibitem[{\citenamefont{Pakter and Levin}(2011)}]{pakter11}
\bibinfo{author}{\bibfnamefont{R.}~\bibnamefont{Pakter}} \bibnamefont{and}
  \bibinfo{author}{\bibfnamefont{Y.}~\bibnamefont{Levin}},
  \bibinfo{journal}{Phys. Rev. Lett.} \textbf{\bibinfo{volume}{106}},
  \bibinfo{pages}{200603} (\bibinfo{year}{2011}).

\bibitem[{\citenamefont{Melchionna}(2007)}]{melchionna07}
\bibinfo{author}{\bibfnamefont{S.}~\bibnamefont{Melchionna}},
  \bibinfo{journal}{J. Chem. Phys.} \textbf{\bibinfo{volume}{127}},
  \bibinfo{pages}{044108} (\bibinfo{year}{2007}).

\bibitem[{\citenamefont{Ram\'{\i}rez-Hern\'andez
  et~al.}(2008)\citenamefont{Ram\'{\i}rez-Hern\'andez, Larralde, and
  Leyvraz}}]{ramirezhernandez09}
\bibinfo{author}{\bibfnamefont{A.}~\bibnamefont{Ram\'{\i}rez-Hern\'andez}},
  \bibinfo{author}{\bibfnamefont{H.}~\bibnamefont{Larralde}}, \bibnamefont{and}
  \bibinfo{author}{\bibfnamefont{F.}~\bibnamefont{Leyvraz}},
  \bibinfo{journal}{Phys. Rev. E} \textbf{\bibinfo{volume}{78}},
  \bibinfo{pages}{061133} (\bibinfo{year}{2008}).

\end{thebibliography}

\end{document}